\documentclass[a4paper,11pt]{article}

\usepackage{pos}
\usepackage{lipsum} %package to generate placeholder text in the following

\usepackage{graphicx}
\usepackage{subcaption}
\usepackage[capitalize]{cleveref}
\usepackage{siunitx}
\usepackage[left]{lineno}

\title{Prospective Sensitivity to Solar Dark Matter using the IceCube Upgrade}
 \ShortTitle{IceCube Upgrade Sensitivity to WIMP annhilation in the core of the Sun}

\author{The IceCube Collaboration \\{\normalsize \normalfont(a complete list of authors can be found at the end of the proceedings)}\\}

\emailAdd{eliot.genton@uclouvain.be}
\emailAdd{jeffrey.lazar@uclouvain.be}
\emailAdd{carguelles@g.harvard.edu}
\emailAdd{gwenhael.dewasseige@uclouvain.be}

\abstract{

While astrophysical observations imply that 85\% of the matter content is unaccounted for, the nature of this dark matter (DM) component remains unknown. Weakly Interacting Massive Particles (WIMPs)—DM particles that interact at or below the weak interaction scale—could naturally explain this missing matter. These interactions with the Standard Model (SM) allow them to be gravitationally captured in celestial bodies like the Sun. Trapped DM in the solar core could subsequently annihilate, producing stable SM particles, of which only neutrinos can escape the Sun’s dense interior. Therefore, an excess of neutrinos originating from the direction of the Sun would serve as evidence of DM. The IceCube Upgrade, a dense infill of the IceCube Neutrino Observatory, will lower the energy threshold and improve sensitivity in the range from 1 to 500~GeV, thereby enhancing IceCube’s ability to detect GeV-scale DM. In this contribution, I present projections of the IceCube Upgrade’s sensitivity to the DM-proton scattering cross section for DM masses between 3~GeV and 500~GeV. These sensitivities position IceCube as the most sensitive indirect detection experiment for DM in the mass range from 3~GeV to 10~TeV. 
\vspace{4mm}

{\bfseries Corresponding authors:}
Jeffrey Lazar$^{1*}$,
Eliot Genton$^{1 2*}$, 
Gwenhaël de Wasseige$^{1}$, 
Carlos Argüelles$^{2}$\\
{$^{1}$ \itshape Center for Cosmology, Particle Physics and Phenomenology, UCLouvain}\\
{$^{2}$ \itshape Laboratory of Particle Physics and Cosmology, Harvard}\\[4mm]
$^*$ Presenter
}

\ConferenceLogo{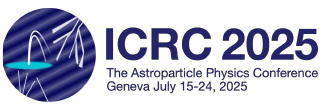}

\FullConference{39th International Cosmic Ray Conference (ICRC2025)\\
 15–24 July 2025\\
Geneva, Switzerland\\}

\begin{document}

\maketitle

\section{Introduction}
\label{sec:introduction}
Neutrinos originating from the Sun have historically provided key insights into both astrophysics and particle physics. 
Observations of low-energy solar neutrinos confirmed the Sun’s energy production via nuclear fusion and revealed the phenomenon of neutrino oscillations, providing evidence for nonzero neutrino mass \cite{, SNO:2001kpb}. 
While solar neutrino studies at MeV energies are well established, the GeV-scale regime remains relatively unexplored and offers new potential for discovery.

Beyond the Standard Model (SM), high-energy neutrino fluxes may arise from the annihilation of dark matter (DM) accumulated in the solar core. Low-mass DM, a well-motivated DM candidate, could become gravitationally captured by the Sun following scatterings with solar nuclei. Over time, an excess of DM may accumulate and annihilate into Standard Model particles, among which only neutrinos can escape from the dense solar environment. These neutrinos offer a unique avenue for indirect detection of dark matter, complementary to terrestrial direct detection experiments \cite{Schumann:2019eaa}.

\section{The IceCube Upgrade}
\label{sec:Icecube_Upgrade}

The IceCube Neutrino Observatory, a cubic-kilometer-scale neutrino telescope located at the geographic South Pole, is uniquely positioned to search for high-energy neutrinos from DM annihilation in the Sun ~\cite{IceCube:2016zyt}. 
It consists of 5,160 Digital Optical Modules (DOMs) deployed at depths between \SI{1,450}{m} and \SI{2,450}{m} along 86 vertical strings, to reduce the contamination from muons produced by the interactions of cosmic rays in the earths atmosphere.
IceCube is designed primarily to detect high-energy neutrinos from astrophysical sources, with optimal sensitivity in the energy range from approximately \SI{100}{\GeV} to several~\si{PeV}. 
Neutrinos are detected indirectly through the Cherenkov radiation emitted by charged secondary leptons, produced in neutrino interactions with the Antarctic ice, that traverse the detector at velocities exceeding the local speed of light. 
This Cherenkov light is captured by the DOMs, enabling the reconstruction of the energy, direction, and flavor of the incident neutrinos.

\begin{figure}[htbp]
    \centering
    \includegraphics[width=0.7\linewidth]{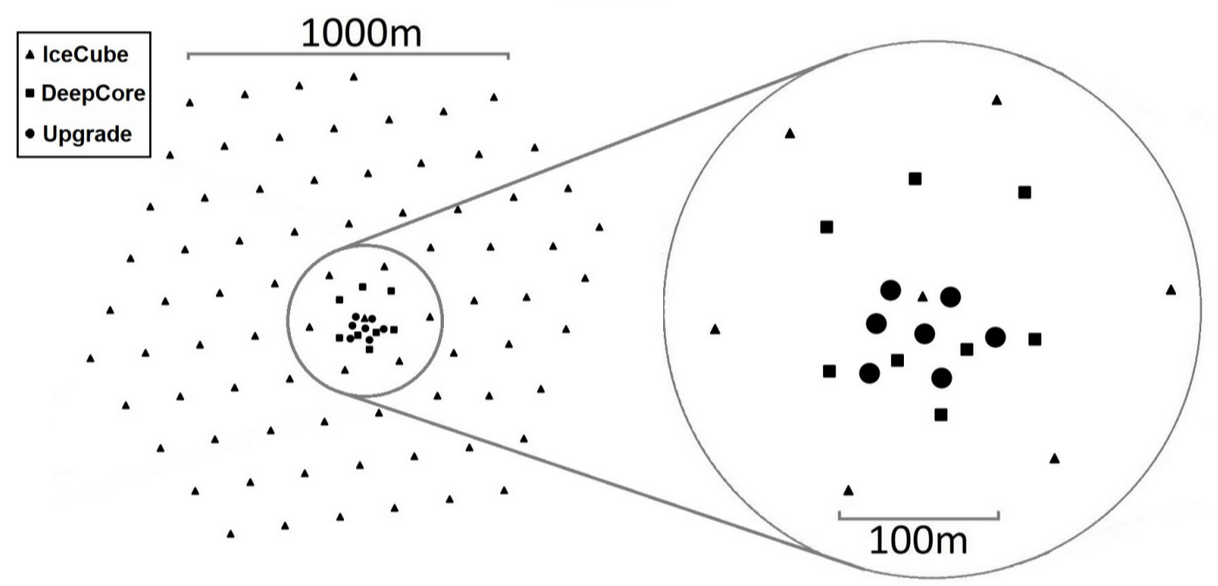}
    \caption{\textbf{\textit{Top-down View of IceCube Detector.}}
    Each point marks a string of optical modules, with symbols indicating sub-detectors. The enlarged view highlights the denser DeepCore and Upgrade regions.
    }
    \label{fig:upgrade}
\end{figure}

\begin{figure}[htbp]
    \centering
    \begin{subfigure}[t]{0.48\textwidth}
        \centering
        \includegraphics[width=\linewidth]{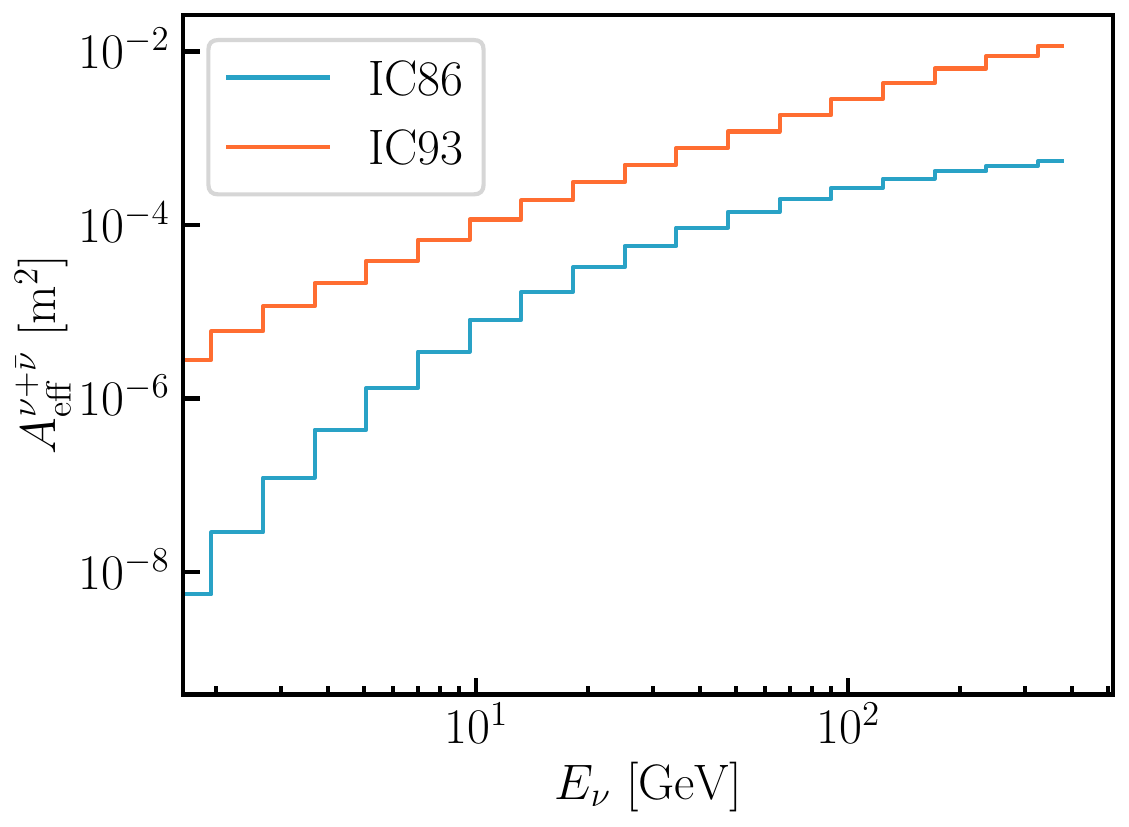}
        \caption{Effective area}
        \label{fig:subfig_angres}
    \end{subfigure}
    \hfill
    \begin{subfigure}[t]{0.48\textwidth}
        \centering
        \includegraphics[width=\linewidth]{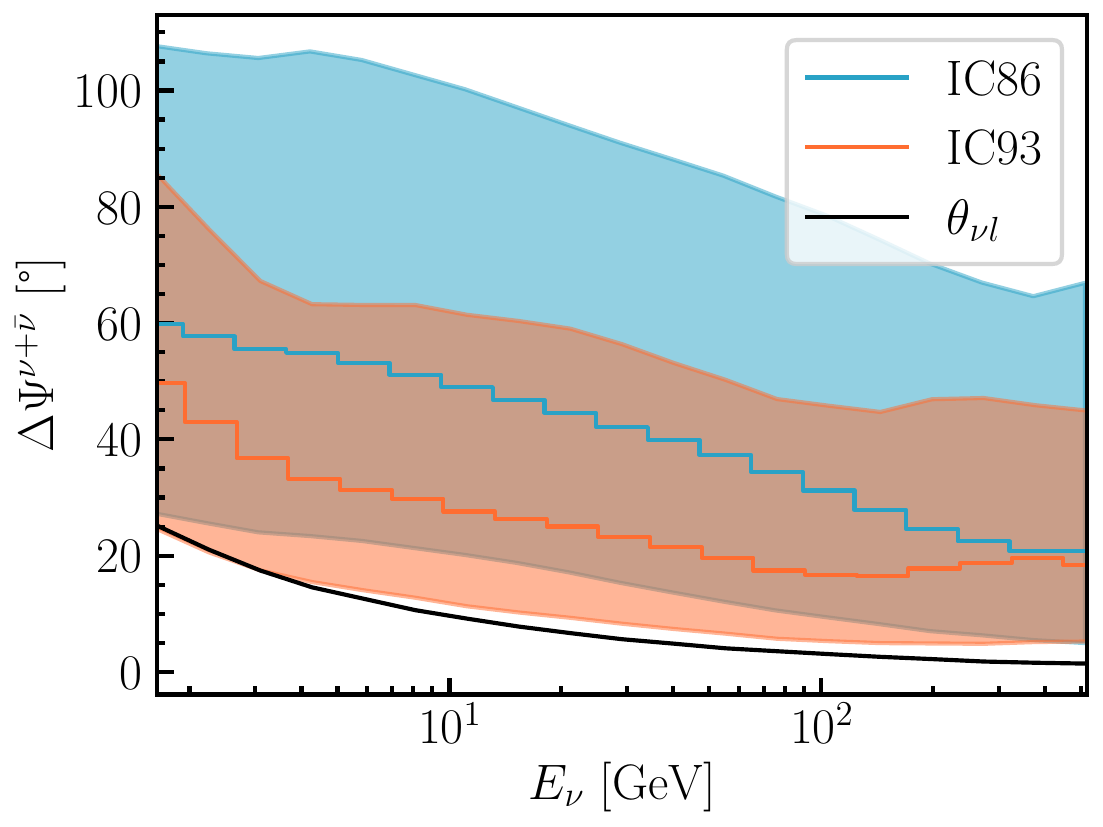}
        \caption{Angular Resolution}
        \label{fig:subfig_effarea}
    \end{subfigure}
    \caption{\textbf{\textit{Detector response comparison between Upgrade and DeepCore.}}  
    Left: effective area as a function of neutrino energy.
    Right: angular resolution (median angle between true and reconstructed direction) as a function of energy; the band shows the $1\sigma$ containment region. The black line indicates median angular separation between the neutrino and lepton in charged-current interactions from the simulation dataset processed  with an IC93 configuration.}
    \label{fig:detector_response}
\end{figure}

The DeepCore subarray, located in the central region of IceCube, features reduced vertical and horizontal spacing between DOMs, resulting in greater sensitivity to faint Cherenkov light and enabling IceCube to detect neutrinos with energies as low as $\mathcal{O}(\mathrm{GeV})$.
This current configuration as of July 2025 is designated as "IC86". 
Looking ahead, the forthcoming IceCube Upgrade will reinforce DeepCore by introducing seven new strings equipped with advanced multi-PMT DOMs, thereby increasing the detector’s DOM density and expanding its fiducial volume. 
This future IceCube Upgrade configuration, designated as "IC93", will push the energy threshold down to $\mathcal{O}(\mathrm{1 GeV})$ and the multi-PMT structure of the added DOMs will also help substantially refine the reconstruction of neutrino direction and energy \cite{Ishihara:2019aao}.
The layout structure of the different mentioned arrays is given in \cref{fig:upgrade}. \cref{fig:subfig_angres} and \cref{fig:subfig_effarea} compare two key reconstruction parameters—angular resolution and effective area—for both IC86 and upgraded IceCube IC94 configurations. 
It should be noted that the sample used to construct the IC86 distributions, which was employed in previous IceCube solar dark matter analyses, was derived from a dataset originally optimized for oscillation studies rather than astronomy.
This selection excludes events with light detected outside the DeepCore region, thereby suppressing higher energy events.

In IceCube, observed neutrino interactions are typically categorized according to their \emph{dominant event topology}: either track-like or cascade-like (shower-like). 
Most events display a single topology, but some can exhibit both track and cascade features simultaneously, depending on the underlying interaction processes. 
Track-like events originate mainly from charged-current $\nu_\mu$ interactions, producing long-range muons that enable precise angular reconstruction. 
Cascade-like events can result from charged-current $\nu_e$ and $\nu_\tau$ interactions, as well as from neutral-current interactions of all neutrino flavors. 
These cascades deposit energy in a more spherically distributed pattern, making directional reconstruction inherently more challenging. 
While track-like events offer superior angular resolution and are ideal for point-source analyses (such as searches for neutrinos from the Sun), cascade-like events provide complementary information through their distinct energy and timing signatures. 

\section{Analysis Method}
\label{sec:Analysis_Method}
DMs captured in massive bodies like the Sun or Earth through scattering with nuclei can accumulate and eventually annihilate, producing a flux of neutrinos \cite{JUNGMAN1996195}. 
The time evolution of the number of DMs, \( N(t) \), in such a body is governed by:
\[
\frac{dN}{dt} = C - C_A N^2 -C_E N,
\]
where \( C \) is the capture rate and \( C_A \) is related to the annihilation cross section and the spatial distribution of DMs within the body and \( C_E \) is the evaporation rate.
DM evaporation, i.e., re-ejection from the Sun  due to rare up-scattering, becomes significant for \( m_\chi \lesssim \SI{3.7}{\GeV} \).
For larger masses considered in this study this process is negligible~\cite{Gould:1987ju}. The annihilation rate is therefore given by:    
\[
\Gamma_A = \frac{1}{2} C_A N^2.
\]
Assuming a constant capture rate and no significant evaporation effects, the solution yields the time-dependent annihilation rate:
\[
\Gamma_A(t) = \frac{1}{2} C \tanh^2\left(\frac{t}{\tau}\right),
\quad \text{where} \quad \tau = \frac{1}{\sqrt{C C_A}}.
\]
If the age of the solar system (\( t \sim 4.5 \) Gyr) is much larger than \( \tau \), equilibrium is reached and the annihilation rate simplifies to \( \Gamma_A \approx \frac{1}{2}C \), meaning the flux of resulting neutrinos is directly proportional to the capture rate. Importantly this capture rate parameter is proportional to the cross section for DM-proton elastic scattering interaction $\sigma_{\chi p}$. For example for an axial-vector (spin-dependant) interaction between DM and the proton constituting the sun, one can write the following expression for the capture rate:

\[
C_{\mathrm{SD}}^{\odot} = (1.3 \times 10^{25}\ \mathrm{s}^{-1})\, \rho_{0.3} \sigma_{\chi p}^{SD} \frac{S(m_{\chi}/m_{\mathrm{H}})}{(m_{\chi}/(1~\mathrm{GeV}))} \overline{v}_{270},
\]
where $\sigma_{\chi p}^{SD}$ is the spin-dependant DM-proton elastic scattering cross section in units of $10^{-40}~\mathrm{cm}^2$, $\overline{v}_{270}$ is the dark matter velocity dispersion in units of $270~\mathrm{km}~\mathrm{s}^{-1}$, and $\rho_{0.3}$ is the local halo mass density in units of $0.3~\mathrm{GeV}~\mathrm{cm}^{-3}$. 
$S(m_\chi/m_{\mathrm{H}})$ is the kinematic suppression factor, $m_\chi$ is the WIMP mass, and $m_{\mathrm{H}}$ is the mass of the hydrogen atom \cite{JUNGMAN1996195}.
WIMPs accumulated in the core of the sun can then annihilate into unstable SM particles which themselves could decay into neutrinos. The direct product of the annhilation is called an annhilation channel e.g $\chi \chi \rightarrow b \bar{b}$.
Therefore, by searching for neutrinos from the direction of the Sun, IceCube can probe the DM-proton scattering cross section, $\sigma_{\chi p}$, providing complementary and, in some parameter regions, competitive constraints to direct detection experiments.

\begin{figure}
    \centering
    \includegraphics[width=0.6\linewidth]{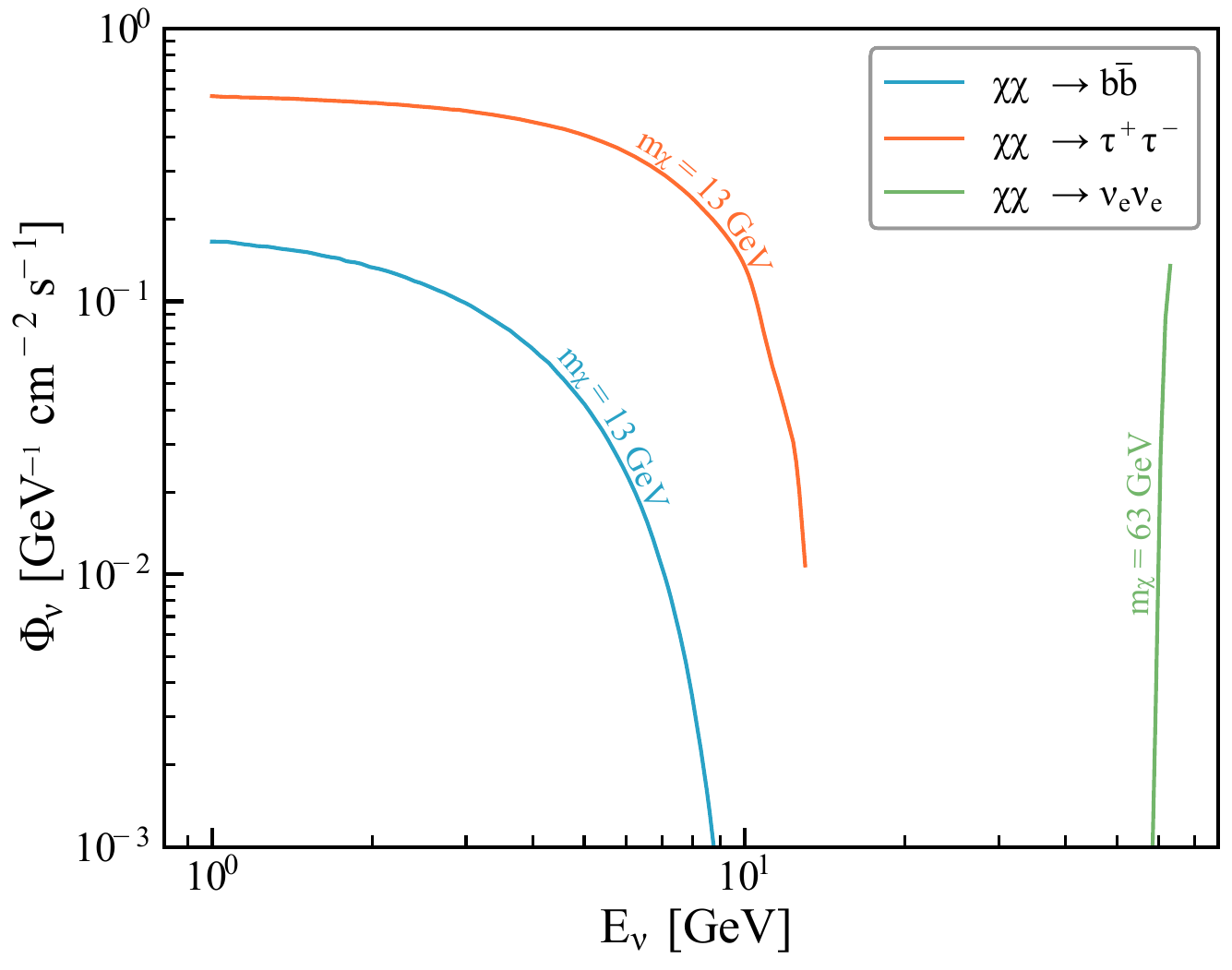}
    \caption{\textbf{\textit{Neutrino flux for different annihilation channels}}
    The line indicates the neutrino flux at the detector for different initial DM mass and for a nominal $\sigma_{\chi p} = \SI{1e-40}{\cm\squared}$
    }
    \label{fig:BSM-Flux}
\end{figure}

\begin{figure}[htbp]
    \centering
    \includegraphics[width=1.0\linewidth]{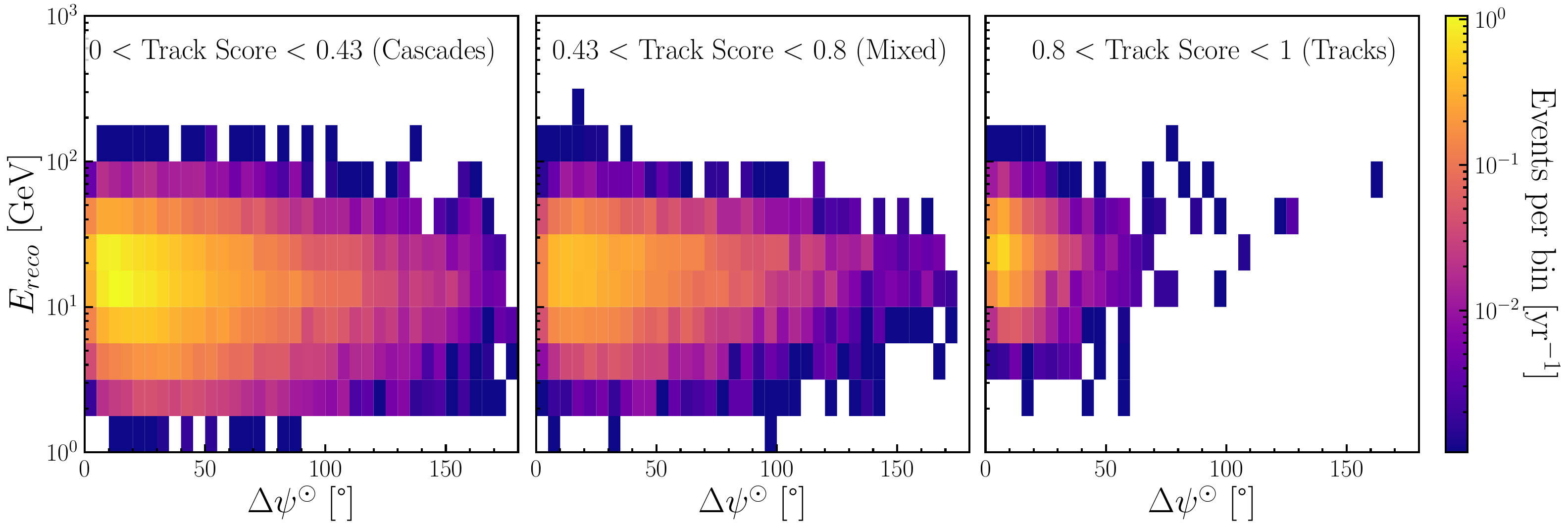}
    \caption{\textbf{\textit{Example signal distribution for solar DM.}} Simulated for DM with mass $m_{\chi} = 63\,\mathrm{GeV}$ and scattering cross-section $\sigma_{\chi p} = \SI{1e-40}{\cm\squared}$, annihilating to $b\bar{b}$ from the solar core. The distribution is binned in reconstructed energy, angular separation from the Sun, and track score.}
    \label{fig:signal_distribution_solar}
\end{figure}

\begin{figure}[htbp]
    \centering
    \includegraphics[width=1.0\linewidth]{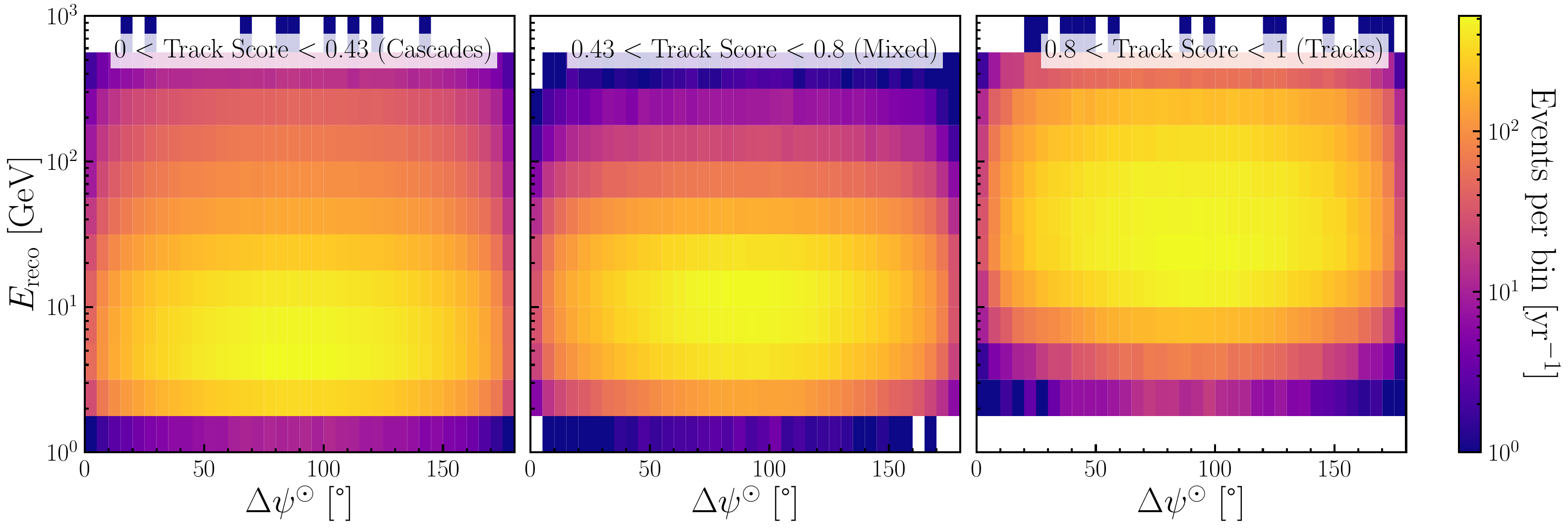}
    \caption{\textbf{\textit{Background distribution for solar analysis.}} Contributions include solar and conventional atmospheric neutrinos and atmospheric muons.
    Noise events are excluded at final selection level.}
    \label{fig:background_distribution_solar}
\end{figure}

To perform this analysis, Monte Carlo simulations are used to construct three-dimensional templates for signal and background, binned in reconstructed energy, angular separation from the Sun, and track score—a classifier distinguishing $\nu_\mu$ charged-current events from other topologies.
Signal events are time-sampled within a three-year window starting January 2026, with Sun positions computed using IceCube software interfacing with the Positional Astronomy Library~\cite{Jenness2013}.
Events are selected if their true direction lies within the solar disk, considered to have a radius of \SI{695700}{\km} and weighted by the product of their MC weight and neutrinos from DM anhilation flux from the \texttt{$\chi$aro$\nu$} tool ~\cite{refId0,Liu_2020}.
Three representative WIMP annihilation channels are considered: $b\bar{b}$, which produces hadronic final states and results in a soft neutrino energy spectrum; $\tau^{+}\tau^{-}$, a leptonic channel yielding a comparatively hard spectrum; and $\nu\bar{\nu}$, which generates a monochromatic neutrino line from two-body decay. The resulting neutrino flux spectra for selected dark matter masses are shown in~\cref{fig:BSM-Flux}.
This process is repeated to yield a smoothed distribution with a reference cross-section $\sigma_{\chi p} = \SI{1e-40}{\cm\squared}$. An example distribution is shown in \cref{fig:signal_distribution_solar}.
As illustrated, events with lower track scores, corresponding to more cascade-like topologies, tend to exhibit poorer angular reconstruction, which is reflected in the broader spread of opening angles to the sun within the distribution.
Background contributions in this analysis include atmospheric muons, conventional atmospheric neutrinos, and solar atmospheric neutrinos. The solar atmospheric neutrino template is constructed identically to the signal template, ensuring consistent treatment of solar directional effects. The atmospheric muon and neutrino backgrounds are generated using IC93 simulation datasets, with event rates weighted according to their respective flux models (GaisserH4a + SIBYLL,IPhonda2014)~\cite{Gaisser:2012zz,Fletcher:1994bd,wren2018thesis}.
To simulate an isotropic background distribution, each muon and neutrino event is assigned a randomized right ascension uniformly distributed in the range \([0, 2\pi)\), along with a random timestamp sampled within the analysis time window which is then used to calculate the opening angle with the Sun. The total background distribution is obtained by summing all three individual components which is shown in Figure \ref{fig:background_distribution_solar}.

The sensitivity in this analysis is estimated using trials within a binned Poisson likelihood framework. The likelihood function is defined as:
\begin{equation}
\mathcal{L}(\boldsymbol{\theta} \,|\, \mathbf{n}) = \prod_{i} \frac{e^{-\mu_i(\boldsymbol{\theta})} \, \mu_i(\boldsymbol{\theta})^{n_i}}{n_i!},
\label{eq:likelihood}
\end{equation}
where $\mathbf{n} = \{n_i\}$ denotes the set of observed event counts in each bin $i$, and $\mu_i(\boldsymbol{\theta})$ is the expected number of events in bin $i$ given the model parameters $\boldsymbol{\theta}$. The parameter vector $\boldsymbol{\theta} = (\alpha_{\chi}, \alpha_{\text{bg}})$ represents the normalization parameters for the potential signal, $\alpha_{\chi}$ (DM-induced solar neutrinos), and for the total background, $\alpha_{\text{bg}}$ (including contributions from muons, solar, and conventional atmospheric neutrinos). The expected event count \(\mu_i(\boldsymbol{\theta})\) in each bin is modeled as a linear combination of the template distributions:
\begin{equation}
\mu_i(\boldsymbol{\theta}) = \alpha_{\chi} \, \mu_{\chi,i} + \alpha_{\text{bg}} \, \mu_{\text{bg},i},
\label{eq:model}
\end{equation}
where $\mu_{\chi,i}$ denotes the expected number of signal events in bin $i$, and $\mu_{\text{bg},i}$ denotes the expected number of background events in bin $i$. The data are binned in $\log_{10}$(reconstructed energy) and in opening angle, with separate bins defined for the track score.  This formulation allows the use of binned Poisson statistics to compare the observed data to the model, and evaluate sensitivity by testing different signal strengths \(\alpha_{\chi}\) through a hypothesis test. The background test statistic (TS) is defined as:
\begin{equation}
\mathrm{TS_{bg}} = -2 \left[ 
\log \mathcal{L}(\hat{\alpha}_{\chi}, \hat{\alpha}_{\text{bg}} \mid \mathbf{n}) 
- 
\log \mathcal{L}(\alpha_{\chi}=0, \check{\alpha}_{\text{bg}},\mid \mathbf{n}) 
\right],
\label{eq:TS}
\end{equation}
where $\check{\alpha}_x$, are best-fit parameters fitting the model including signal from DM anhilation $\theta(\alpha_\chi,\alpha_{bg}=1)$ onto a data set $\mathbf{n}$ sampled from the background only model $\theta(\alpha_\chi=0,\alpha_{bg}=1)$. Due to non-negativity constraints on fit parameters, the $\mathrm{TS}_{bg}$ follows a truncated $\chi^2$ distribution with one degree of freedom: 50\% of cases yield TS = 0, and 90\% fall below $\mathrm{TS}_{bg}$ = 1.64. This threshold is adopted for setting sensitivity targets. Sensitivity is computed as the median signal normalization that results in a $\mathrm{TS}_{sig}$ $=$ 1.64 across many trials, providing a robust measure for signal detectability under the given analysis strategy.

\begin{equation}
\mathrm{TS_{sig}(\alpha_\chi)} = -2 \left[ 
\log \mathcal{L}(\alpha_{\chi}, \hat{\alpha}_{\text{bg}} \mid \mathbf{n}) 
- 
\log \mathcal{L}(\hat{\alpha}_\chi, \hat{\alpha}_{\text{bg}},\mid \mathbf{n}) 
\right],
\label{eq:TS}
\end{equation}

\section{Results}
\label{sec:results}
\begin{figure}[t]
    \centering
    \includegraphics[width=0.7\linewidth]{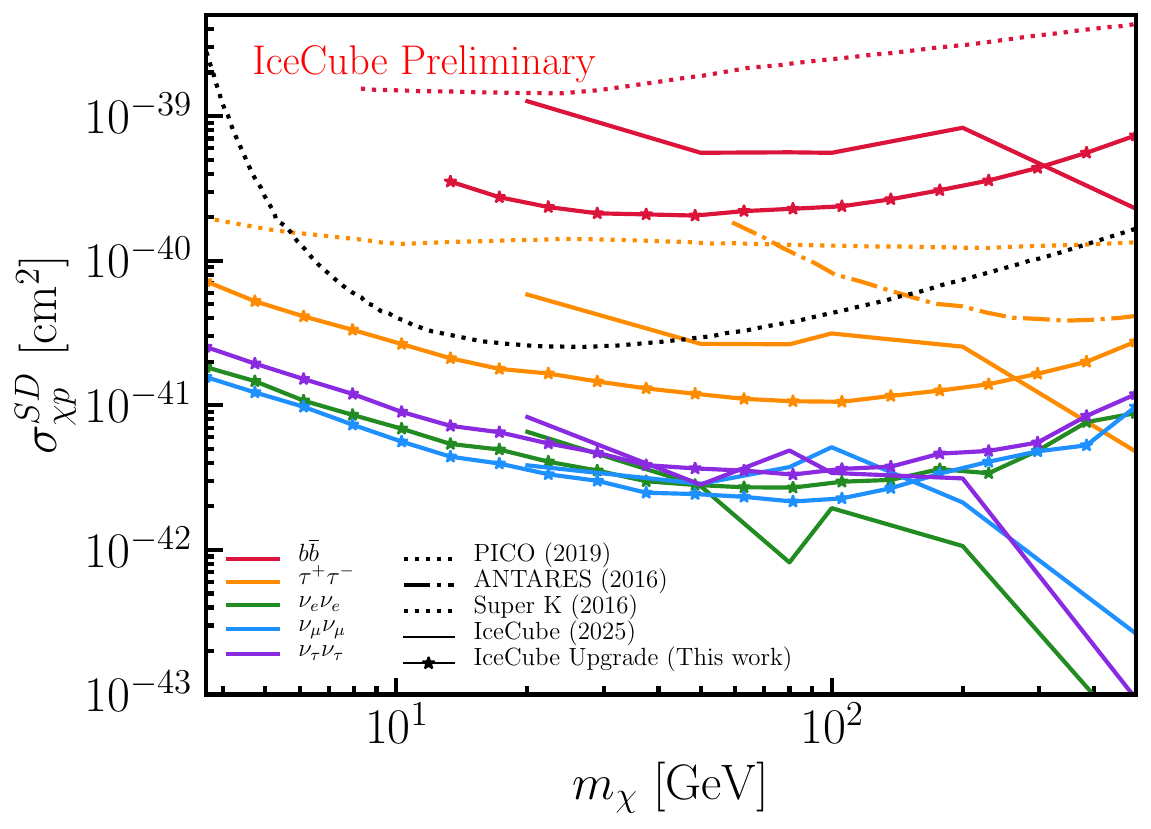}
    \caption{
        \textbf{\textit{Sensitivity to DM-proton scattering $\sigma_{\chi p}$ with three years of data.}} 
        These projected limits from the IceCube Upgrade would extend IceCube's reach down to \SI{3}{\GeV}, offering complementarity to the lowest masses probed by direct detection experiments. 
        Furthermore, the IceCube Upgrade is expected to deliver world-leading limits on DM-proton scattering cross sections below $\mathcal{O}$(100) GeV. 
        The projected sensitivities are compared to existing limits from \cite{Adri_n_Mart_nez_2016, kamiokandecollaboration2015searchneutrinosannihilationcaptured,Amole_2019}.
    }
    \label{fig:Solar_DM_sensitivity}
\end{figure}
The sensitivity results, obtained following the methodology in \cref{sec:Analysis_Method}, are summarized in \cref{fig:Solar_DM_sensitivity} and compared with recent results from leading dark matter searches~\cite{Adri_n_Mart_nez_2016,kamiokandecollaboration2015searchneutrinosannihilationcaptured,Amole_2019}. The IceCube Upgrade is projected to achieve the best sensitivity to low-mass DM annihilation up to $\sim 210$\,GeV for the $b\bar{b}$ and $\tau^+\tau^-$ channels, and up to $\sim 20$\,GeV for neutrino final states. The minimum DM mass tested is limited by different factors depending on the channel: for $\nu$ and $\tau^+\tau^-$, the constraint arises from the assumption of negligible DM evaporation, valid only above $\sim 3.7$\,GeV~\cite{Gould:1987ju}; for $b\bar{b}$, it stems from the use of \texttt{PYTHIA} in the $\chi aro \nu$ package, which does not reliably simulate hadronic decays below $\sim 10$\,GeV ~\cite{Sj_strand_2015}.

\section{Conclusion}
\label{sec:conclusion}

This work presents a sensitivity study for detecting neutrinos from low-mass dark matter annihilation in the Solar Core using the planned IceCube Upgrade. A dedicated template-based analysis was developed to search for excess neutrino events from the solar direction, incorporating both track-like and cascade-like topologies. Template construction utilized full MC simulations and flux models for signal and background components, while sensitivity estimates were derived via binned Poisson likelihood pseudo-experiments. The analysis achieves competitive sensitivity projections with three years of data, extending IceCube’s reach down to DM masses as low as 3.7 GeV and providing leading limits below $\mathcal{O}(100\,\mathrm{GeV})$ for several annihilation channels. The projected performance of the IceCube Upgrade reaffirms its power as a next-generation facility in the search for physics beyond the Standard Model.

% Bibtex references:
\bibliographystyle{ICRC}
\bibliography{references}

% Alternatively, you can include references by hand:
%\begin{thebibliography}{99}
%\bibitem{...}
%
%\end{thebibliography}

\clearpage

%The following list of authors, affiliations and funding agencies will be updated at the day of submission. The following template is a placeholder generated via https://authorlist.icecube.wisc.edu/icecube on May 17, 2025 and will be updated.
\section*{Full Author List: IceCube Collaboration}

\scriptsize
\noindent
R. Abbasi$^{16}$,
M. Ackermann$^{63}$,
J. Adams$^{17}$,
S. K. Agarwalla$^{39,\: {\rm a}}$,
J. A. Aguilar$^{10}$,
M. Ahlers$^{21}$,
J.M. Alameddine$^{22}$,
S. Ali$^{35}$,
N. M. Amin$^{43}$,
K. Andeen$^{41}$
% ...
\\
\\
% ...
$^{10}$ Universit{\'e} Libre de Bruxelles, Science Faculty CP230, B-1050 Brussels, Belgium \\
% ...
$^{16}$ Department of Physics, Loyola University Chicago, Chicago, IL 60660, USA \\
$^{17}$ Dept. of Physics and Astronomy, University of Canterbury, Private Bag 4800, Christchurch, New Zealand \\
% ...
$^{21}$ Niels Bohr Institute, University of Copenhagen, DK-2100 Copenhagen, Denmark \\
$^{22}$ Dept. of Physics, TU Dortmund University, D-44221 Dortmund, Germany \\
% ...
$^{35}$ Dept. of Physics and Astronomy, University of Kansas, Lawrence, KS 66045, USA \\
% ...
$^{39}$ Dept. of Physics and Wisconsin IceCube Particle Astrophysics Center, University of Wisconsin{\textemdash}Madison, Madison, WI 53706, USA \\
% ...
$^{41}$ Department of Physics, Marquette University, Milwaukee, WI 53201, USA \\
% ...
$^{43}$ Bartol Research Institute and Dept. of Physics and Astronomy, University of Delaware, Newark, DE 19716, USA \\
% ...
$^{63}$ Deutsches Elektronen-Synchrotron DESY, Platanenallee 6, D-15738 Zeuthen, Germany \\
$^{\rm a}$ also at Institute of Physics, Sachivalaya Marg, Sainik School Post, Bhubaneswar 751005, India \\
% ...

\subsection*{Acknowledgments}

\noindent
The authors gratefully acknowledge the support from the following agencies and institutions:
USA {\textendash} U.S. National Science Foundation-Office of Polar Programs,
U.S. National Science Foundation-Physics Division,
U.S. National Science Foundation-EPSCoR,
U.S. National Science Foundation-Office of Advanced Cyberinfrastructure,
Wisconsin Alumni Research Foundation,
Center for High Throughput Computing (CHTC) at the University of Wisconsin{\textendash}Madison,
Open Science Grid (OSG),
Partnership to Advance Throughput Computing (PATh),
Advanced Cyberinfrastructure Coordination Ecosystem: Services {\&} Support (ACCESS),
Frontera and Ranch computing project at the Texas Advanced Computing Center,
U.S. Department of Energy-National Energy Research Scientific Computing Center,
Particle astrophysics research computing center at the University of Maryland,
Institute for Cyber-Enabled Research at Michigan State University,
Astroparticle physics computational facility at Marquette University,
NVIDIA Corporation,
and Google Cloud Platform;
Belgium {\textendash} Funds for Scientific Research (FRS-FNRS and FWO),
FWO Odysseus and Big Science programmes,
and Belgian Federal Science Policy Office (Belspo);
Germany {\textendash} Bundesministerium f{\"u}r Forschung, Technologie und Raumfahrt (BMFTR),
Deutsche Forschungsgemeinschaft (DFG),
Helmholtz Alliance for Astroparticle Physics (HAP),
Initiative and Networking Fund of the Helmholtz Association,
Deutsches Elektronen Synchrotron (DESY),
and High Performance Computing cluster of the RWTH Aachen;
Sweden {\textendash} Swedish Research Council,
Swedish Polar Research Secretariat,
Swedish National Infrastructure for Computing (SNIC),
and Knut and Alice Wallenberg Foundation;
European Union {\textendash} EGI Advanced Computing for research;
Australia {\textendash} Australian Research Council;
Canada {\textendash} Natural Sciences and Engineering Research Council of Canada,
Calcul Qu{\'e}bec, Compute Ontario, Canada Foundation for Innovation, WestGrid, and Digital Research Alliance of Canada;
Denmark {\textendash} Villum Fonden, Carlsberg Foundation, and European Commission;
New Zealand {\textendash} Marsden Fund;
Japan {\textendash} Japan Society for Promotion of Science (JSPS)
and Institute for Global Prominent Research (IGPR) of Chiba University;
Korea {\textendash} National Research Foundation of Korea (NRF);
Switzerland {\textendash} Swiss National Science Foundation (SNSF).

\end{document}